# Analyzing Prominent LLMs: An Empirical Study of Performance and Complexity in Solving LeetCode Problems

Everton Guimaraes
*EASER, Eng. Division*
Penn State University
Malvern, USA
ezt157@psu.edu

Nathalia Nascimento
*EASER, Eng. Division*
Penn State University
Malvern, USA
nqm5742@psu.edu

Asish Nelapati
*EASER, Eng. Division*
Penn State University
Malvern, USA
akn5618@psu.edu

Chandan Shivalingaiah
*EASER, Eng. Division*
Penn State University
Malvern, USA
cms9109@psu.edu

*Abstract*—Large Language Models (LLMs) like ChatGPT, Copilot, Gemini, and DeepSeek are transforming software engineering by automating key tasks, including code generation, testing, and debugging. As these models become integral to development workflows, a systematic comparison of their performance is essential for optimizing their use in real-world applications. This study benchmarks these four prominent LLMs on 150 LeetCode problems across easy, medium, and hard difficulties, generating solutions in Java and Python. We evaluate each model based on execution time, memory usage, and algorithmic complexity, revealing significant performance differences. ChatGPT demonstrates consistent efficiency in execution time and memory usage, while Copilot and DeepSeek show variability as task complexity increases. Gemini, although effective on simpler tasks, requires more attempts as problem difficulty rises. Our findings provide actionable insights into each model's strengths and limitations, offering guidance for developers selecting LLMs for specific coding tasks and providing insights on the performance and complexity of GPT-like generated solutions.

## I. INTRODUCTION

The rapid evolution of Generative AI (GenAI), particularly Large Language Models (LLMs), has revolutionized software engineering, bringing transformative advancements to tasks like code generation, testing, and debugging [19], [22]. The advent of tools like ChatGPT has propelled LLMs to considerable popularity, extending their application beyond general purposes to those tailored exclusively for code-related tasks. These specialized LLMs aim to save developers time and effort by streamlining various programming tasks.

Examples of coding-specific models include CodeBERT [11], Codex [10] and PolyCoder [30], which support developers by embedding directly into Integrated Development Environments (IDEs) as code assistants, like GitHub Copilot. Meanwhile, broader LLMs such as ChatGPT, Copilot, Gemini, and DeepSeek have proven adept not only at generating code but also at performing additional tasks, including metric calculation, code complexity analysis, and documentation generation. This versatility has fueled interest in comparing the strengths and limitations of various LLMs for software engineering workflows [7], [19].

Previous studies have demonstrated the effectiveness of individual LLMs across various programming tasks, such as code generation, code completion, and bug detection. Chen et al. [10] showed that Codex excels at generating code from natural language descriptions, while Feng et al. [11] highlighted the proficiency of CodeBERT in code search and generation tasks. Furthermore, Svyatkovskiy et al. [26] found that LLMs can significantly enhance code completion, and Allamanis et al. [1] explored their potential in bug detection. These findings provide a foundation for our research, which aims to compare the performance of different LLMs on a standardized set of algorithmic challenges, focusing on execution time, memory usage, and algorithmic complexity.

This study examines four leading LLMs—ChatGPT, Copilot, Gemini, and DeepSeek—by evaluating their performance on 150 algorithm and data structure problems from the LeetCode platform. To address this gap in comparative analysis, this study analyzes solutions generated in Java and Python, focusing on execution time, memory usage, and algorithmic complexity (both time and space). By exclusively assessing LLM-generated solutions, our research aims to offer a comprehensive comparison, shedding light on each model's performance across varying difficulty levels [14], [21].

The growing interest in role-specific AI tools underscores the value of continual improvements in model reliability, accuracy, and efficiency within software engineering [17], [20]. While many studies explore individual model capabilities, few provide comparative analyses that could guide optimal model selection for diverse engineering tasks. Our study contributes to this knowledge base by assessing both performance and complexity metrics. These insights aim to help developers and job-seekers make more informed decisions about which LLMs are best suited for general software engineering tasks, including code generation and optimization. By evaluating these prominent LLMs on a standardized set of algorithmic challenges, this study advances our understanding of their capabilities and provides a structured framework for integrating

LLMs more effectively into professional software engineering environments.

This paper offers the following key contributions to advancing LLM evaluation in software engineering:

- We conduct a controlled evaluation of four prominent LLMs (ChatGPT, Copilot, Gemini, and DeepSeek) on 150 algorithm and data structure problems from LeetCode, assessing their code generation capabilities across varying levels of problem difficulty.
- We measure each model's performance across execution time, memory usage, and complexity (both time and space) in solutions generated in both Java and Python, offering insights into language-specific model performance.
- We analyze and rank each model's solution complexity, highlighting which models provide the most efficient solutions and how model efficiency shifts with increased problem complexity.

The remainder of this paper is organized as follows. Section II reviews the literature on LLMs in software engineering, focusing on code generation and model evaluation benchmarks. Section III describes our research methodology, including experiment design, dataset selection, and performance metrics. Section IV presents data analysis and findings for each research question. Section V discusses threats to validity and the study's limitations. Finally, Section VI concludes with key insights and future research directions, including extending our methodology to additional LLMs and coding benchmarks.

## II. RELATED WORK

GenAI tools like GitHub Copilot and ChatGPT have rapidly integrated into software engineering workflows, facilitating tasks such as code generation, debugging, and testing [4], [24], [29]. These LLMs assist with activities from code completion to error detection, enhancing productivity but also posing challenges, such as generating plausible yet incorrect code, or 'hallucinations,' which underscore the need for continual refinement and responsible use [24], [27].

### A. LLMs in Code Generation and Their Evaluation

Previous research has investigated various aspects of LLMs for code-related tasks, focusing on the quality and correctness of the code they generate [8], [16], [21]. Studies have examined how developers interact with these models [3], [27] and the prevalence of bugs in the generated code [13]. Several benchmarks have been developed to assess LLMs' capabilities in code generation. Notable benchmarks include HumanEval [9], CoderEval [6], CodeXGLUE [18], and ReCode [2]. These benchmarks provide standardized tasks to evaluate LLM performance on coding challenges. However, we wanted to focus on coding platforms (LeetCode [15], StrataScratch [25]) that are prominently used by developers, engineers, and jobseekers.

### B. Evaluations Using LeetCode Problems

LeetCod [15], primarily a coding competition platform, has been utilized as a dataset to evaluate the capabilities of LLMs on programming tasks. Döderlein et al. [8] measured the performance of Copilot and Codex on LeetCode problems and investigated the effects of changing prompts. Nguyen and Nadi [21] studied GitHub Copilot's code suggestions on LeetCode problems, analyzing the complexities of its generated code. Vasconcelos et al. [28] examined the effects of highlighting the uncertainty of AI-powered code completions using LeetCode problems and Codex.

Nguyen and Nadi [21] also evaluated the correctness and understandability of GitHub Copilot's code in solving LeetCode problems. They used the number of passed test cases and SonarQube, a static code analysis platform, for evaluation. However, they did not consider performance, memory efficiency, or compare the results with human-written solutions.

### C. Performance and Efficiency of LLM-Generated Code

Several techniques have been proposed for automatically enhancing code performance using LLMs [5], [12]. However, to the best of our knowledge, our contribution is the first to investigate the differences in the performance of LLM-generated code across different models on LeetCode problems. Future LLMs adapted with these performance-improving techniques could also be compared using our methodology.

## III. RESEARCH METHODOLOGY

This study investigates the performance of GenAI models in solving coding challenges, focusing on execution time, memory consumption, and complexity (detailed in Section III-B4) for solutions generated in Python and Java. We use problems typically found in technical interviews for software engineering positions to mirror real-world coding scenarios. These problems emphasize algorithmic concepts and efficient problem-solving. LeetCode is a suitable platform for this evaluation due to its extensive repository of categorized problems, standardized format, built-in testing environment, and clear difficulty levels. This structure allows for a systematic assessment of LLM-generated code across various complexities.

### A. Research Questions

This study aims to provide empirical evidence to guide developers in making informed decisions, enhancing both productivity and code quality, by evaluating the execution time, memory consumption, and complexity of code generated in Java and Python. We selected Java and Python for this study due to their prevalence in software engineering and consistent ranking among the top three most popular programming languages according to the TIOBE index[1]. Our research questions are as follows:

**RQ1:** *How do LLMs perform across different problem difficulties in terms of execution time and memory usage?* This question examines LLM capabilities across varying task complexities, essential for developers seeking efficient solutions that scale with real-world problem demands.

---

[1] See the TIOBE index at [https://www.tiobe.com/tiobe-index/](https://www.tiobe.com/tiobe-index/)

**RQ2:** *Is there a significant difference in the efficiency of solutions generated by each LLM in Java versus Python?* By assessing efficiency in Java and Python, we explore language-specific performance, helping developers choose the most suitable LLM for their programming language of choice.

**RQ3:** *Which LLM produces the most optimized solutions in terms of time and space complexity across difficulty levels?* This question identifies the LLM with the most optimized outputs, enabling developers to select models that produce efficient code for performance-sensitive applications.

**RQ4:** *How does problem difficulty impact the accuracy of LLM-generated complexity metrics in Java and Python?* Examining the effect of difficulty on complexity metrics allows us to assess LLM reliability in estimating resource demands, informing developers of each model's strengths and limitations.

### B. Overview of Experimental Design

We conducted a controlled experiment to evaluate the performance of several LLMs in generating code solutions. The experiment was designed to systematically assess model performance across problem difficulties and programming languages, focusing on the metrics discussed in Section III-B4.

*1) Data Collection:* We curated a dataset of 150 programming problems from LeetCode, equally divided by difficulty levels: easy, medium, and hard, with 50 problems in each category. The problems focus on algorithms and data structures, offering solutions with varied performance characteristics, enabling us to observe and compare the outputs generated by different LLMs. LeetCode was chosen for its structured focus on programming challenges, its clear problem categorization, and its active developer community. Its emphasis on time and space efficiency through Big-O complexity aligns with our objectives of assessing LLM performance in terms of execution time, memory usage, and complexity optimization.

LeetCode's extensive library also provides a balanced difficulty spectrum, which is ideal for systematically evaluating LLM capabilities. Our methodology is inspired by the approach used in [7], though we applied it to a different set of LeetCode questions. Solutions were generated in both Java and Python to analyze any potential language-specific performance differences.

*2) Selection of Large Language Models:* We evaluated the following widely used LLMs, such as ChatGPT, Copilot, and Gemini (see Table I). These models were chosen based on their accessibility and popularity among developers. We included **DeepSeek**, an open-source model recognized for strong performance in code generation tasks according to [23], to compare mainstream proprietary LLMs with a top-performing open-source alternative.

The GenAIs/LLMs listed in Table I represent some of the most advanced LLMs currently available, each known for its distinct capabilities and parameter scale. ChatGPT (GPT-4o) and Microsoft Copilot (GPT-4), both built on OpenAI's GPT-4 architecture, are estimated to contain around 1.8 trillion parameters, enabling them to perform complex coding, conversational, and reasoning tasks. Google Gemini (Version 1.0), another leading model with over 1 trillion parameters, is designed for multimodal inputs, processing both text and image data with advanced reasoning capabilities. DeepSeek (Coder 2.5), an open-source alternative with approximately 238 billion parameters, leverages a mixture of experts to optimize task-specific responses, providing a more accessible yet powerful solution in code generation and general AI tasks.

| Generative AI Model | LLM Version | Estimated Parameters |
|---|---|---|
| ChatGPT | GPT-4o | Approximately 1.8 trillion |
| Microsoft Copilot | GPT-4 | Approximately 1.8 trillion |
| Google Gemini | Version 1.0 | Over 1 trillion |
| DeepSeek | Coder 2.5 | 238 billion |

TABLE I
GenAI Models and LLM Versions

*3) Prompt Design and Experimental Procedure:* Our controlled experiment included a total of four LLMs for general purposes. Each model was guided to produce code solutions that could be systematically evaluated using the metrics outlined in Section III-B4. As part of the experimental procedure, each LLM generated solutions for the curated 150 problems in both Java and Python. This dual-language approach enabled an analysis of performance differences between the languages for each model. Each model was allowed up to three attempts per problem per language to produce a correct and optimized solution, enabling a robust comparison across models and languages. This study used a structured prompt to guide LLMs in generating consistent and effective code solutions. Designed to simulate a real-world coding task, the prompt directed the LLM to act as a software developer, standardizing responses across models to enable fair comparison. Below, we summarize the key prompt components.

The prompt structure is illustrated in Figure 1. The prompt instructs the LLM to "act as a software developer," establishing a role-based context that primes the model for a problem-solving approach with an emphasis on accuracy and code quality. It includes a problem description, examples, and constraints, helping the LLM recognize input structures and align responses with standard coding problem formats. Specifying Java as the output language ensures consistency across responses, reducing variations that could arise from different default languages in the models. Additionally, the LLM is instructed to calculate the Big O complexity, which prompts it to assess the efficiency of its solution—an essential factor in our performance evaluation. Finally, specifying `Solution.java` as the file name mirrors software development practices, enabling consistent organization of outputs and facilitating automated evaluation. Note that the prompt design minimizes ambiguity and encourages standardized outputs across LLMs. The structured instructions promote both functional correctness and efficiency, aligning with real-world coding practices and simplifying automated testing and evaluation.

*4) Evaluation Metrics:* As part of the experimental procedure, the generated solutions were submitted to LeetCode's

Fig. 1. Prompt Design Overview

```
# Start of the Prompt

Query:
Act as a software developer and provide a solution for the problem
described below. Note that the problem has a description, examples,
and constraints. When you finish implementing the solution, you need
to calculate the complexity using the Big O notation. The source code
must be developed in Java, and the generated file should be named
Solution.

Problem Description:
Given an input string s and a pattern p, implement regular expression
matching with support for '.' and '*' where:

'.' Matches any single character.
'*' Matches zero or more of the preceding element.
The matching should cover the entire input string (not partial).

Example 1:
Input: s = "aa", p = "a"
Output: false
Explanation: "a" does not match the entire string "aa".

Example 2:
Input: s = "aa", p = "a*"
Output: true
Explanation: '*' means zero or more of the preceding element, 'a'.
Therefore, by repeating 'a' once, it becomes "aa".

Constraints:
1 <= s.length <= 20
1 <= p.length <= 20

# End of Prompt.
```

online judge system, where automated tests evaluated their correctness, execution time, and memory usage. This standardized evaluation ensured consistency in assessing each solution's performance metrics. We collected the following metrics for each generated solution:

- **Execution Time**: The time taken by the solution to run, as reported by LeetCode's automated testing system.
- **Memory Usage**: The amount of memory consumed during the execution of the solution.
- **Complexity Analysis**: The time and space complexity of the solution, derived from analyzing the generated code in Big-O notation.

### C. Replication Package

A replication package will be made accessible upon the paper's acceptance. It will contain all artifacts related to this study, including results, prompts, generated source code, datasets, and Jupyter Notebook.

## IV. DATA ANALYSIS AND RESULTS

This section summarizes the key results from our experiment and addresses the research questions. By examining the performance of LLMs across different problem difficulties, we can identify the strengths and weaknesses of each model and determine which are best suited for specific tasks

### A. RQ1: Performance Across Problem Difficulties:

To assess the performance of various LLMs across different leet code problem difficulties (easy, medium, hard), we segmented the data by difficulty level and evaluated the execution time and memory usage for each LLM within each category. This segmentation allowed us to analyze the mean execution time and memory usage within each difficulty category. This analysis provides insights into the models' handling of computationally simple versus complex tasks, allowing for a comparative view of their efficiency in time and space management.

TABLE II
DESCRIPTIVE STATISTICS OF EXECUTION TIME AND MEMORY USAGE BY
DIFFICULTY LEVEL AND LLM

| LLM | Difficulty | Mean Execution Time (ms) | Mean Memory Usage (MB) |
|---|---|---|---|
| Copilot | Easy | 21.67 ($\pm$ 7.28) | 42.35 ($\pm$ 6.82) |
| Copilot | Medium | 134.54 ($\pm$ 22.87) | 46.15 ($\pm$ 7.96) |
| Copilot | Hard | 342.98 ($\pm$ 59.34) | 53.74 ($\pm$ 9.21) |
| DeepSeek | Easy | 18.52 ($\pm$ 6.17) | 43.87 ($\pm$ 6.74) |
| DeepSeek | Medium | 112.87 ($\pm$ 20.44) | 48.52 ($\pm$ 8.13) |
| DeepSeek | Hard | 298.73 ($\pm$ 52.63) | 54.89 ($\pm$ 8.65) |
| ChatGPT | Easy | 13.47 ($\pm$ 5.28) | 40.78 ($\pm$ 5.98) |
| ChatGPT | Medium | 29.43 ($\pm$ 10.58) | 43.02 ($\pm$ 7.23) |
| ChatGPT | Hard | 47.56 ($\pm$ 15.34) | 46.89 ($\pm$ 7.78) |
| Gemini | Easy | 20.34 ($\pm$ 6.12) | 43.62 ($\pm$ 6.34) |
| Gemini | Medium | 102.23 ($\pm$ 21.93) | 47.15 ($\pm$ 8.11) |
| Gemini | Hard | 254.43 ($\pm$ 48.52) | 52.67 ($\pm$ 9.47) |

*1) Execution Time Distribution by Difficulty and LLM*: The results of the ANOVA and Kruskal-Wallis tests show statistically significant differences in execution time between LLMs at the hard level (F=5.07, $p < 0.01$), while results for easy and medium levels were not significant ($p > 0.05$). The Kruskal-Wallis test revealed significant differences for easy tasks (H=24.47, $p < 0.001$), but not for medium or hard levels.

ChatGPT consistently demonstrated lower execution times across all difficulty levels, averaging 13.47 ms ($\pm$ 5.28) for easy, 29.43 ms ($\pm$ 10.58) for medium, and 47.56 ms ($\pm$ 15.34) for hard tasks, outperforming other models in time efficiency. Both Copilot and DeepSeek showed increased execution times with higher task difficulty, with Copilot averaging 342.98 ms ($\pm$ 59.34) at the hard level. Gemini's performance aligned with DeepSeek and Copilot, while ChatGPT remained more stable across complexities. These findings suggest ChatGPT's advantage in handling complex tasks with resilience and efficiency, whereas Copilot, DeepSeek, and Gemini exhibit greater execution time and variability as complexity rises.

*2) Memory Usage Distribution by Difficulty and LLM*: For memory usage, ANOVA and Kruskal-Wallis tests indicated no significant differences at any difficulty level (p¿0.05 across all tests). However, Kruskal-Wallis tests identified notable differences at the easy (H=20.50, $p < 0.001$) and medium (H=17.63, $p < 0.01$) levels, suggesting that memory usage varies primarily in simpler tasks. ChatGPT also showed consistent performance in memory usage across all levels, averaging 40.78 MB ($\pm$ 5.98) at the easy level and gradually increasing to 46.89 MB ($\pm$ 7.78) at the hard level. By comparison, Copilot, DeepSeek, and Gemini exhibited increased memory usage and broader variability at the hard level, suggesting greater demands on memory as task complexity rises.

This pattern suggests that while ChatGPT manages memory more efficiently, Copilot, DeepSeek, and Gemini may need more memory resources to handle complex tasks, especially at higher difficulty levels.

*3) Insights from the Box Plots and Outlier Analysis:* The box plots (see Fig 2 and Fig 3) offer deeper insights into the variability and central tendencies of execution time and memory usage across LLMs by difficulty level.

**Execution Time and Memory Usage Analysis.** Distinct patterns in execution time and memory usage emerge across the LLMs, highlighting ChatGPT's efficiency and stability at all difficulty levels. For execution time, ChatGPT demonstrates relatively low and stable times across tasks, reflecting its optimized handling of both simple and complex inputs. At the easy and medium levels, Copilot, DeepSeek, and Gemini exhibit minimal variability with close median times, indicating efficient performance for straightforward tasks. However, at the hard level, execution time variability increases significantly for Copilot and DeepSeek, with higher medians and broader interquartile ranges, suggesting challenges in managing complex tasks and occasional inefficiencies. In terms of memory usage, ChatGPT consistently shows a lower median across all levels, especially at the easy and medium levels. Copilot, DeepSeek, and Gemini display similar memory usage at the easy level but with slightly broader distributions, indicating minor inefficiencies even in simpler tasks. As task complexity rises to the medium level, all models show a modest increase in memory usage variability, though ChatGPT maintains a more contained distribution, suggesting effective memory management. At the hard level, Copilot, DeepSeek, and Gemini require significantly more memory with wider distributions, reflecting increased resource demands and less consistent memory management as solution complexity grows.

**Outliers and Comparative Efficiency**. The box plots reveal key outliers in Copilot's and DeepSeek's execution times and memory usage, especially at higher difficulty levels. For instance, some of Copilot's solutions require significantly more memory than the median, possibly due to intensive data handling and nested operations aimed at boosting execution speed. DeepSeek also shows several outliers in both execution time and memory usage, suggesting occasional inefficiencies in its ability to handle the computational load for complex tasks. ChatGPT, on the other hand, displays minimal outliers across all metrics, reinforcing its stability in both time and memory efficiency. Overall, ChatGPT emerges as the most time- and memory-efficient model across difficulty levels, particularly at the hard level where performance consistency is critical. While efficient for simpler tasks, Copilot, DeepSeek, and Gemini show increased execution time and memory usage variability as task complexity escalates. These findings suggest that ChatGPT may be a more suitable choice for applications requiring scalable performance across varying complexities, whereas the other models may be more resource-intensive and less consistent in high-demand scenarios.

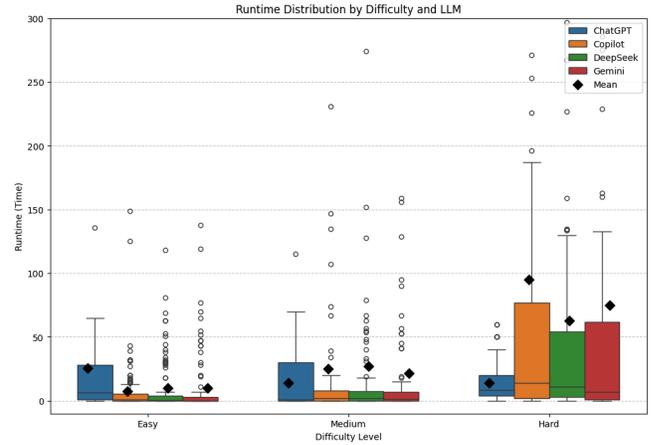

Fig. 2. Runtime Distribution by Difficulty and LLM

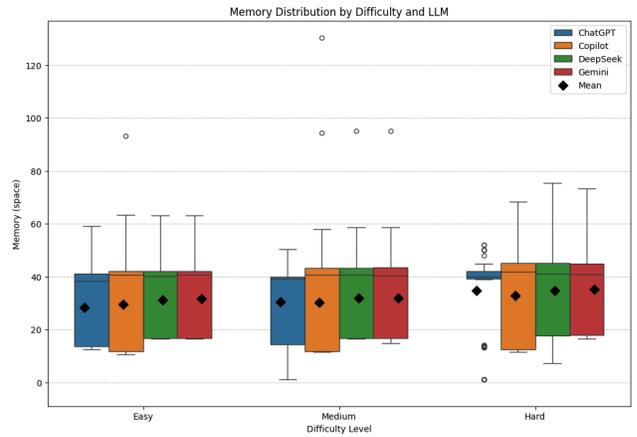

Fig. 3. Memory Distribution by Difficulty and LLM

## B. RQ2: Efficiency in Java vs. Python

We compared the efficiency of solutions produced by each LLM in Java versus Python

*1) Descriptive Statistics and Initial Analysis:* To investigate the efficiency of solutions generated by each LLM (Copilot, DeepSeek, ChatGPT, and Gemini) in Java versus Python, we examine both execution time and memory usage. Table III provides the descriptive statistics, including mean and standard deviation for each metric across languages.

TABLE III
DESCRIPTIVE STATISTICS OF EXECUTION TIME AND MEMORY USAGE BY LANGUAGE AND LLM

| LLM | Language | Mean Execution Time (ms) | Mean Memory Usage (MB) |
|---|---|---|---|
| Copilot | Java | 19.91 ($\pm$ 3.12) | 45.35 ($\pm$ 7.29) |
| Copilot | Python | 228.66 ($\pm$ 14.85) | 15.83 ($\pm$ 2.21) |
| DeepSeek | Java | 11.87 ($\pm$ 2.46) | 45.34 ($\pm$ 6.91) |
| DeepSeek | Python | 149.97 ($\pm$ 13.24) | 19.55 ($\pm$ 3.89) |
| ChatGPT | Java | 6.85 ($\pm$ 1.95) | 46.76 ($\pm$ 8.42) |
| ChatGPT | Python | 29.63 ($\pm$ 4.83) | 15.87 ($\pm$ 2.13) |
| Gemini | Java | 13.07 ($\pm$ 3.18) | 44.75 ($\pm$ 5.87) |
| Gemini | Python | 18.52 ($\pm$ 2.94) | 19.38 ($\pm$ 3.54) |

From Table III, it is evident that Java solutions generally exhibit faster execution times across all LLMs, while Python solutions tend to be more memory-efficient. These initial descriptive statistics provide a foundational understanding of the differences between Java and Python in terms of computational efficiency. To assess the significance of these differences, we performed paired t-tests for execution time and memory usage across Java and Python solutions generated by each LLM. The following sections provide detailed comparative insights based on the statistical analysis.

*2) Execution Time Comparison*: The results show a statistically significant difference in execution time between Java and Python for each LLM (p-value < 0.05). For Copilot, Java solutions had an average execution time of 19.91 ms (± 3.12), which is substantially faster than Python's average execution time of 228.66 ms (± 14.85). This considerable difference suggests that Java solutions generated by Copilot are notably more efficient in terms of speed. Similarly, for DeepSeek, Java solutions were found to be faster, with an average of 11.87 ms (± 2.46) compared to Python's 149.97 ms (± 13.24). Although the gap is smaller than with Copilot, Java remains significantly faster than Python in this case as well.

Moreover, ChatGPT also demonstrated an advantage for Java in terms of execution efficiency, with Java solutions averaging 6.85 ms (± 1.95) compared to Python's 29.63 ms (± 4.83). Finally, for Gemini, Java solutions averaged 13.07 ms (± 3.18) while Python solutions averaged 18.52 ms (± 2.94), showing a smaller but consistent advantage in execution speed for Java across different models.

Overall, Java solutions tend to be faster across all LLMs, indicating an inherent advantage in execution time, possibly due to structural efficiencies in Java solutions or language characteristics. Python's slower execution times could stem from language overheads or different solution structures generated by the LLMs.

*3) Memory Usage Comparison*: Memory usage also reveals statistically significant differences, with Python generally consuming less memory than Java. For Copilot, Java solutions required an average memory usage of 45.35 MB (± 7.29), whereas Python solutions used only 15.83 MB (± 2.21). This indicates that, despite Java's faster execution times, it consumes significantly more memory than Python. DeepSeek exhibits similar trends, with Java solutions averaging 45.34 MB (± 6.91) in memory usage compared to Python's 19.55 MB (± 3.89). ChatGPT further reinforces this pattern, as Java solutions consumed an average of 46.76 MB (± 8.42), while Python solutions required only 15.87 MB (± 2.13), highlighting Python's consistent memory efficiency advantage. Lastly, Gemini also aligns with this trend, with Java solutions averaging 44.75 MB (± 5.87) compared to Python's 19.38 MB (± 3.54). Across all models, Java consistently uses more memory than Python, suggesting a clear trade-off between execution speed and memory usage in these language implementations. This consistent pattern suggests that Python solutions tend to be more memory-efficient across all LLMs, likely due to streamlined data handling or concise solution structures.

*4) Insights from the Box Plots and Outlier Analysis*: The box plots (see Figures 4 and 5) reveal additional insights and outliers. The box plots reveal specific execution time outliers for Python solutions, especially in Copilot and DeepSeek. In Copilot, some Python solutions exhibit execution times well above 300 ms, far exceeding the median. For example, a few Python solutions for complex problems display prolonged execution times, suggesting that Copilot occasionally generates less optimized Python code structures that struggle with computational efficiency. Similarly, DeepSeek shows Python outliers with execution times approaching 200 ms, underscoring variability in Python's performance, especially for computationally demanding tasks.

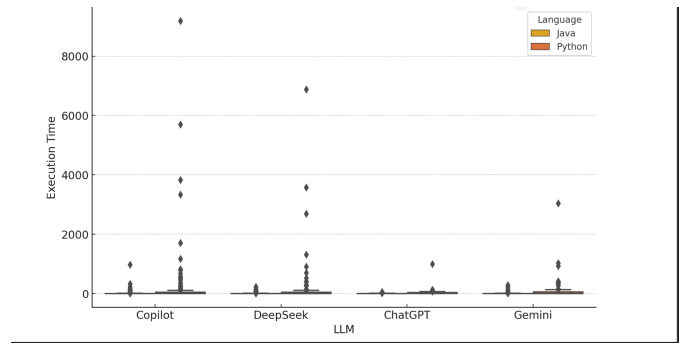

Fig. 4. Comparison of Execution Time by Language for Each LLM

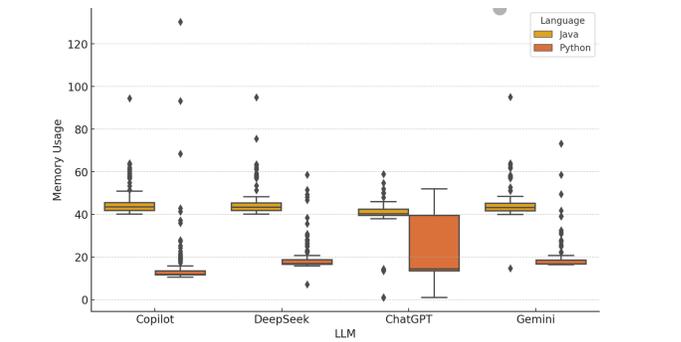

Fig. 5. Comparison of Memory Usage by Language for Each LLM

In terms of memory usage, Java solutions consistently show higher values across all LLMs, with extreme cases appearing in ChatGPT and Copilot. For instance, Copilot has a few Java solutions requiring over 60 MB of memory—significantly more than the median—likely due to complex or nested data structures used to enhance execution speed. ChatGPT also displays Java memory usage outliers, with certain solutions surpassing 55 MB, indicating that specific tasks may drive Java's memory allocation requirements higher than usual.

Overall, these outliers illustrate a trade-off in efficiency between Java and Python solutions across all LLMs. Java solutions are generally faster but tend to consume more memory, while Python solutions are slower but typically more memory-efficient. This pattern highlights distinct language characteris-

tics and the LLMs' differing approaches to problem-solving in each language

*C. RQ3: Optimization of Solutions*

To address this research question we assessed the time and space complexity of the solutions generated by each LLM across varying difficulties. First, we standardized the time complexities generated by the LLMs and ranked them giving a numerical value to each complexity. The lower the value, the better the complexity. Then, averages were calculated, and significance was tested using the Friedman test. Results were represented using stacked bar charts to compare the proportion of solutions at different complexity levels.

| Complexity | Numerical Rank |
|---|---|
| $O(1)$ | 1 |
| $O(\log n)$ | 2 |
| $O(\log n^2)$ | 2 |
| $O(n)$ | 3 |
| $O(n \log n)$ | 4 |
| $O(n^2)$ | 5 |
| $O(n^2 \log n)$ | 5 |
| $O(n^3)$ | 6 |
| $O(n^{81})$ | 7 |
| $O(m^n)$ | 8 |
| $O(4^n)$ | 8 |
| $O(2^n)$ | 8 |
| $O(3^n)$ | 8 |
| $O(n!)$ | 9 |

TABLE IV
COMPLEXITY RANKINGS

*1) Preliminary Analysis:* The dataset provides a structured comparison of time and space complexity for solutions generated by various LLMs across different difficulty levels (Easy, Medium, and Hard). To quantify optimization, each complexity level was assigned a numerical rank, with lower ranks indicating more efficient performance. For each LLM, average complexity ranks were calculated, and the Friedman test was applied to assess the significance of differences in time and space complexities among the models.

*2) Time Complexity Distribution by LLM and Difficulty:*
The data reveals distinct time complexity patterns across LLMs at Easy, Medium, and Hard difficulty levels. ChatGPT and Gemini exhibit relatively consistent distributions of lower time complexity ranks, with ChatGPT notably maintaining low time complexity as difficulty increases, indicating efficient time management across tasks. In contrast, Copilot and DeepSeek display broader spreads, particularly at medium and hard levels, suggesting less consistency and higher time complexity for more challenging problems.

ChatGPT shows a strong concentration in lower time complexity ranks (Ranks 1-4) across all difficulties, signifying superior optimization for time efficiency. Conversely, Copilot and DeepSeek exhibit a broader range into higher complexity ranks (Ranks 7-9) as difficulty increases, indicating challenges in maintaining efficiency under complex conditions. Overall, ChatGPT emerges as the most time-optimized model, with Gemini performing comparably well though occasionally exhibiting higher complexity at the medium level. Copilot and DeepSeek appear less optimized, tending towards higher time complexities with increased difficulty. The Friedman test (statistic = 8.20, $p = 0.042$) confirms a statistically significant difference in time complexity across LLMs, with a p-value below 0.05 indicating that some models are notably better optimized than others.

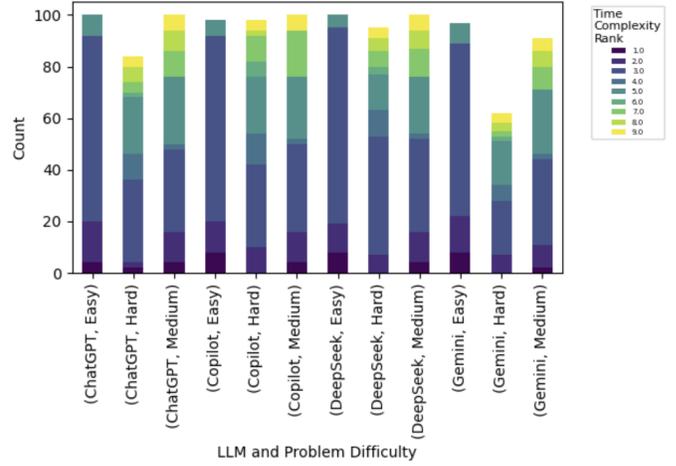

Fig. 6. Time Complexity Distribution by LLM and Difficulty

*3) Space Complexity Distribution by LLM and Difficulty:*
The space complexity analysis shows distinct clustering for each model across difficulty levels, contrasting with the varied time complexity distributions. ChatGPT demonstrates low space complexity ranks on easy tasks, indicative of efficient space usage, but spreads into higher ranks on medium and hard tasks, reflecting increased space demands with rising difficulty. DeepSeek generally maintains middle-range space complexity ranks, without optimization for the lowest ranks.

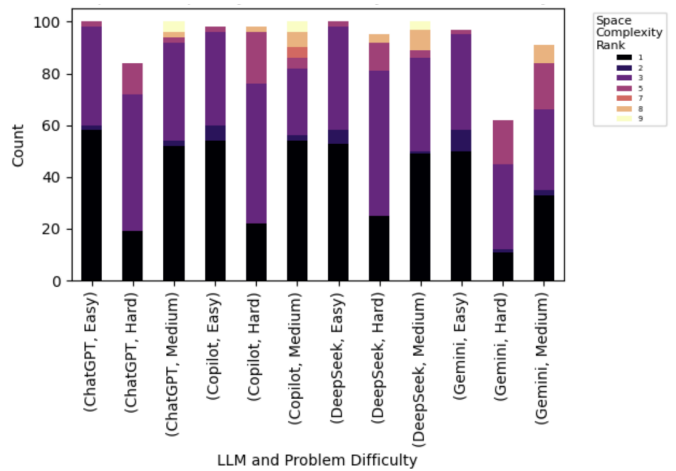

Fig. 7. Space Complexity Distribution by LLM and Difficulty

Higher space complexity levels (Ranks 7-9) are more frequent for Copilot and DeepSeek as task difficulty increases,

suggesting less efficiency in space management under complex conditions. ChatGPT demonstrates effective space optimization on simpler tasks by maintaining lower space complexity ranks. Gemini and Copilot show more varied distributions, with Gemini mixed across difficulty levels and Copilot shifting towards higher ranks on difficult tasks. DeepSeek appears less optimized for space complexity, with broader spreads into higher ranks under complex scenarios. The Friedman test for space complexity (statistic = 6.60, $p = 0.086$) indicates no statistically significant difference at the 0.05 level, suggesting less variation in space complexity performance among these models compared to time complexity.

In conclusion, ChatGPT consistently produces time-optimized solutions across varying difficulties, with lower time complexity ranks and significant differences supported by Friedman test results. While Gemini performs well in both time and space complexity, its effectiveness diminishes at higher difficulty levels. Copilot and DeepSeek show broader distributions into higher complexity ranks as difficulty escalates, indicating less consistent optimization.

### D. RQ4 - Impact of Problem Difficulty on Complexity

We analyzed the accuracy of LLM-generated complexity metrics by problem difficulty and programming language using mixed-effects models. This approach accounted for both fixed effects (e.g., problem difficulty, programming language) and random effects (e.g., individual problems). Accuracy trends were visualized with bar charts and box plots to illustrate how problem difficulty influenced the correctness of the complexity assessment.

| Level | LLM | $O(1)$ | $O(2^n)$ | $O(3^n)$ | $O(4^n)$ | $O(\log n)$ | $O(m^n)$ | $O(n)$ | $O(n^2)$ |
|---|---|---|---|---|---|---|---|---|---|
| E | GPT | 58 | 0 | 0 | 0 | 2 | 0 | 38 | 2 |
| E | CP | 54 | 0 | 0 | 0 | 6 | 0 | 38 | 2 |
| E | DS | 53 | 0 | 0 | 0 | 5 | 0 | 40 | 2 |
| E | GG | 50 | 0 | 0 | 0 | 8 | 0 | 37 | 2 |
| H | GPT | 19 | 0 | 0 | 0 | 0 | 0 | 53 | 12 |
| H | CP | 22 | 2 | 0 | 0 | 0 | 0 | 54 | 20 |
| H | DS | 25 | 3 | 0 | 0 | 0 | 0 | 56 | 11 |
| H | GG | 11 | 0 | 0 | 0 | 1 | 0 | 33 | 17 |
| M | GPT | 52 | 2 | 0 | 0 | 2 | 0 | 38 | 2 |
| M | CP | 54 | 6 | 0 | 0 | 2 | 4 | 26 | 4 |
| M | DS | 49 | 5 | 0 | 0 | 3 | 1 | 36 | 3 |
| M | GG | 33 | 4 | 1 | 2 | 2 | 0 | 31 | 18 |

TABLE V
PERFORMANCE COMPARISON OF LLMs ACROSS COMPLEXITY CLASSES

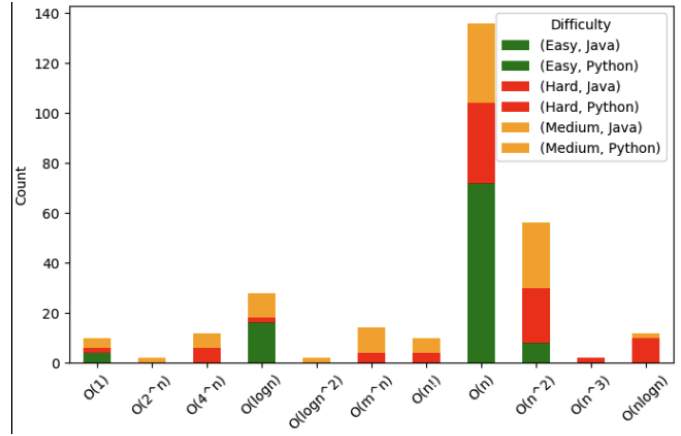

Fig. 8. Standardized Time Complexity for ChatGPT

*1) Preliminary Analysis:* Table V compares the performance of four LLMs—GPT (ChatGPT), CP (Copilot), DS (DeepSeek), and GG (Gemini)—across complexity classes at Easy (E), Medium (M), and Hard (H) difficulty levels. Each model's effectiveness is evaluated for time complexities ranging from $O(1)$ to $O(n^2)$. GPT excels with 58 instances in $O(1)$ for Easy tasks but struggles with higher complexities like $O(n!)$ and $O(n^2)$. Copilot shows consistent strength in Easy tasks with more variable results in Medium and Hard levels. DeepSeek and Gemini also vary in performance by complexity class, with Gemini notably strong in handling higher complexities at the Hard level, particularly for $O(n^2)$. This analysis reveals each LLM's strengths and limitations as task complexity increases.

*2) Comparative Analysis of Time Complexity Distributions Across LLMs:* The stacked bar charts reveal distinct patterns in their handling of standardized time complexities across different difficulty levels and programming languages. While the visual representation offers a clear breakdown of counts by complexity class, language, and difficulty, deeper insights can be derived by analyzing the underlying trends and variations across the models.

ChatGPT shows a strong concentration in the $O(n)$ complexity class, particularly for easy Java problems, suggesting a preference for linear solutions in simpler tasks (see Figure 8). However, it also exhibits higher complexities like $O(n^2)$ for hard Python problems, indicating challenges in optimizing complex Python tasks. In contrast, Copilot displays a balanced distribution across $O(n)$ and $O(n^2)$, with fewer hard Java problems but a notable presence in medium Python tasks at $O(n)$. This hints at a slight optimization edge in Python for medium tasks and fewer instances of high-complexity solutions like $O(2^n)$ in hard problems. Similarly, DeepSeek aligns with Copilot in its $O(n)$ concentration for easy Java tasks, showing efficient handling of simpler problems. However, like ChatGPT, it struggles with hard Python tasks, as indicated by its $O(n^2)$ representation, suggesting limitations in optimizing complex Python solutions.

Finally, Gemini displays a unique pattern with a relatively lower overall count but a strong preference for easy Java problems in the $O(n)$ class (see Figure 9. Gemini also shows fewer instances of high-complexity solutions ($O(2^n)$, $O(3^n)$), suggesting that it may avoid overly complex algorithms even for harder problems. This could indicate either an inherent bias towards simpler solutions or better optimization strategies compared to other models.

*3) Comparative Analysis of Space Complexity Trends Across LLMs:* The Gemini model shows a strong preference for constant space complexity ($O(1)$), with many solutions for both easy and medium problems in Java and Python falling into this category. Python solutions for harder prob-lems also contribute to $O(1)$, though less prominently. The model exhibits a balanced spread across $O(n)$ space complexity, spanning all difficulty levels and languages, especially medium-difficulty problems in Java. This pattern implies that Gemini explores various strategies for different difficulties, while generally favoring memory efficiency.

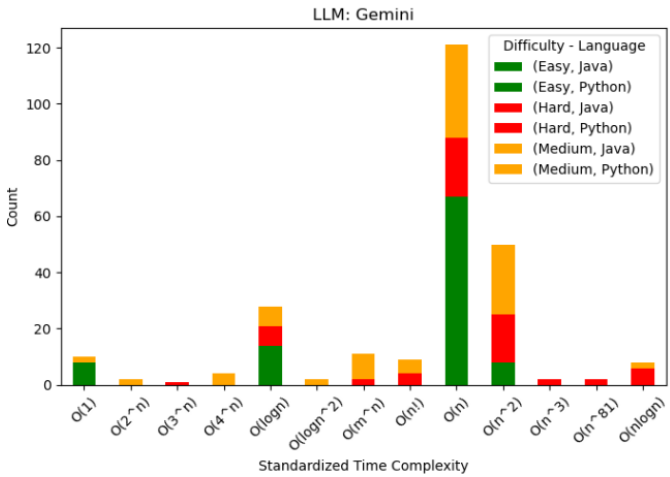

Fig. 9. Standardized Time Complexity for Gemini

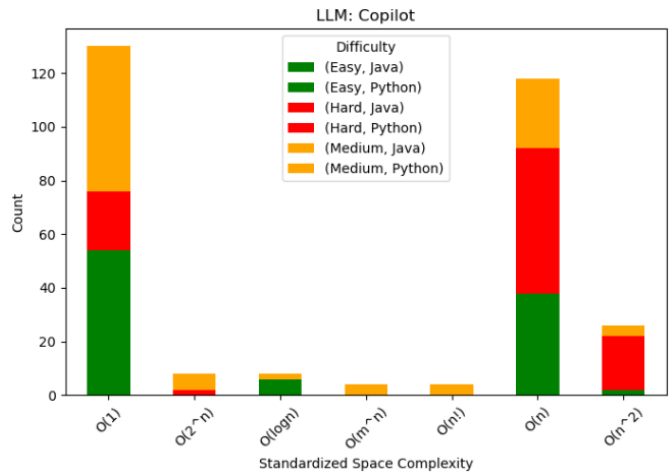

Fig. 10. Standardized Space Complexity for Copilot

Copilot displays a similar trend, concentrating most solutions in $O(1)$, especially for easy problems in both Java and Python. However, its distribution in $O(n)$ is more prominent, particularly for hard problems in Python and Java (see Figure 10). This suggests that Copilot may adopt slightly more complex memory usage for harder problems, while showing minimal engagement with higher space complexities like $O(2^n)$ or $O(\log n)$, indicating a focus on memory optimization.

In turn, ChatGPT also favors $O(1)$ space complexity, dominated by easy Java and Python problems. Similar to Copilot, ChatGPT has a significant spread in the $O(n)$ category, especially for hard problems in both languages. What distinguishes ChatGPT is its even distribution across medium-level problems in both Java and Python within $O(n)$, suggesting a more varied approach in tackling medium problems compared to the other models.

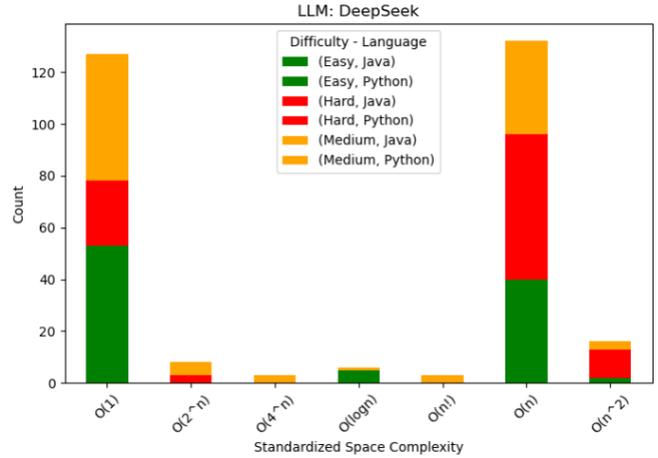

Fig. 11. Standardized Space Complexity for DeepSeek

Finally, DeepSeek generally follows the pattern of favoring $O(1)$, with a notable spread across $O(n)$, particularly for hard problems in Python and Java. Interestingly, DeepSeek shows a slight increase in higher space complexities like $O(2^n)$, though these cases are rare (see Figure 11). This suggests that while DeepSeek leans toward memory-efficient solutions, it occasionally explores more complex algorithms requiring greater memory.

## V. Discussion

This section examines each Large Language Model's (LLM) success rates and efficiency across problem difficulties, highlighting the reliability of GPT, Copilot, Gemini, and DeepSeek in solving easy, medium, and hard tasks. Through success rate and attempt trends, we identify model strengths, limitations, and opportunities for improvement, guiding effective model selection for diverse coding challenges.

### A. Success Rate by Problem Difficulty Level

The analysis reveals that GPT and Copilot consistently perform well, achieving a 100% success rate on both easy and medium problems. Gemini follows closely, with slightly lower success rates, particularly in harder problems (68%). Moreover, the analysis shows that DeepSeek now performs comparably to other LLMs, achieving a 100% success rate on easy and medium problems and a 96% success rate on hard problems. This change brings DeepSeek closer to GPT, Copilot, and Gemini in terms of reliability, particularly across easier and moderate problem levels.

Figure 12 highlights the comparative success rates across difficulty levels for each LLM, clearly showing their relative reliability in solving problems. It can be used to support discussions on model strengths and potential areas for improvement in handling varying problem complexities.

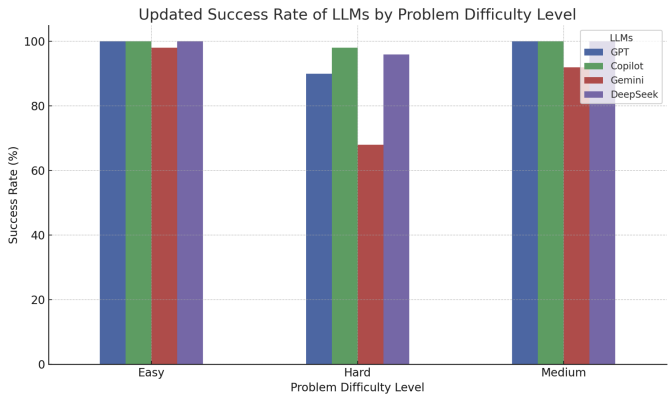

Fig. 12. Success Rate of LLMs by Problem Difficulty Level

## B. Trend of Average Number of Attempts per Difficulty Level

This section examines each LLM's efficiency in generating correct solutions, as shown by the average number of attempts across problem difficulties. Figure 13 illustrates the average number of attempts each LLM model requires to generate a correct solution. The analysis shows that GPT, Copilot, and DeepSeek exhibit high efficiency on easy and medium problems, maintaining close to a single attempt on average. This consistency extends to hard problems, where they require only slightly more attempts, indicating robust performance and reliable success across various problem complexities. In contrast, Gemini displays an upward trend in attempts with increasing problem difficulty, averaging 1.58 attempts on hard problems. This pattern suggests that Gemini may face greater challenges in generating correct solutions for complex tasks, requiring additional iterations compared to other models.

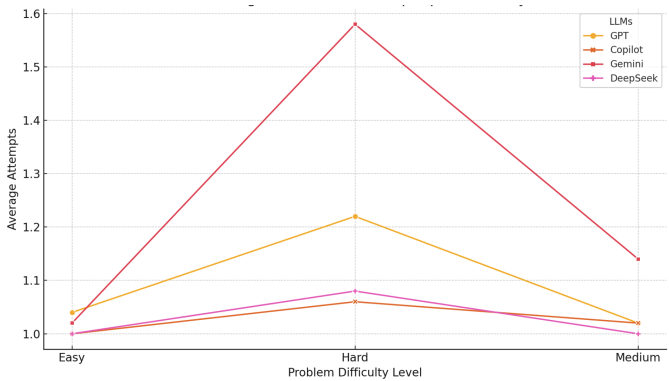

Fig. 13. Trend of Average Number of Attempts per Difficulty Level

The low average attempts for GPT, Copilot, and DeepSeek across difficulty levels make them strong choices for environments where rapid and consistent solution generation is crucial. These models demonstrate efficiency that benefits both developers and teams working across a range of coding challenges. Gemini's increased attempts for harder problems highlight opportunities for refinement, particularly in handling complex tasks.

## VI. THREATS TO VALIDITY

This section outlines potential threats to the validity of our study and our efforts to mitigate them, ensuring reliable and applicable findings across coding scenarios.

### A. Internal Validity

Internal validity concerns the extent to which results can be attributed to the LLMs rather than external factors. We mitigate **model versioning inconsistencies** by using fixed versions for each LLM and ensure **prompt consistency** with standardized prompts across models. To reduce **execution environment variability**, we perform multiple test iterations per problem, averaging the results to lessen server fluctuations on LeetCode. These strategies help ensure that performance reflects each LLM's capabilities.

### B. External Validity

External validity addresses the generalizability of our findings. We use a **balanced set of LeetCode problems** across difficulties to approximate real-world challenges. By including both proprietary and open-source LLMs, we aim to enhance **model representativeness**, though future work may benefit from adding newer models and datasets like CodeSignal. Additionally, focusing on **Java and Python** captures high-level scripting and object-oriented paradigms; future studies could expand this to languages such as C++ and JavaScript.

### C. Construct Validity

Construct validity ensures we are accurately measuring the intended performance metrics. Our key metrics—**execution time**, **memory usage**, and **algorithmic complexity**—are reliable indicators of coding efficiency. However, **Big-O complexity** may not fully capture real-world performance under variable conditions, and focusing solely on **execution time and memory** may overlook aspects like **code readability** and **adherence to coding standards**. Future analyses incorporating these additional constructs would provide a more comprehensive view of LLM performance.

## VII. FINAL REMARKS AND FUTURE WORK

Our study investigated the performance of four leading LLMs—ChatGPT, Copilot, Gemini, and DeepSeek—on code generation tasks, analyzing execution time, memory usage, and complexity. We found that ChatGPT exhibits remarkable consistency in both execution time and memory efficiency, even at higher complexity levels, challenging previous research that emphasizes LLM variability [21]. Our analysis also revealed a distinct trade-off: Java solutions generally yielded faster execution times, while Python solutions demonstrated greater memory efficiency. This nuanced finding highlights the importance of considering both execution speed and memory usage in AI-assisted software development.

While some might assume that LLMs consistently produce optimized solutions, our results show variability in complexity levels, particularly for Copilot and DeepSeek as task difficulty increases. This underscores the need for further research into LLM optimization strategies. Future work will analyze additional open-source LLMs and investigate other problem categories using alternative benchmarking platforms. We also intend to assess solution correctness and similarity by comparing LLM-generated solutions to standard platform-provided ones. This research has significant implications for the development and application of LLMs in software engineering, paving the way for more efficient AI-assisted software development.

## REFERENCES

[1] ALLAMANIS, M., BROCKSCHMIDT, M., AND KHADEMI, M. Learning to represent edits. In *International Conference on Learning Representations*.

[2] ANONYMOUS. Recode: Benchmarking language models for code repair and generation. https://github.com/example/recode, 2023. ReCode Benchmark.

[3] BARKE, S., JAMES, M. B., AND POLIKARPOVA, N. Grounded copilot: How programmers interact with code-generating models, 2022.

[4] BENDER, E. M., GEBRU, T., MCMILLAN-MAJOR, A., AND SHMITCHELL, S. On the dangers of stochastic parrots: Can language models be too big? In *Proceedings of the 2021 ACM Conference on Fairness, Accountability, and Transparency* (2021), pp. 610–623.

[5] CHEN, Z., FANG, S., AND MONPERRUS, M. Supersonic: Learning to generate source code optimizations in c/c++. *IEEE Transactions on Software Engineering* (2024), 1–17.

[6] CODEREVAL. Codereval: A benchmark for evaluating code generation models, 2022. Benchmark for evaluating code generation models.

[7] COIGNION, T., QUINTON, C., AND ROUVOY, R. A performance study of llm-generated code on leetcode. In *28th International Conference on Evaluation and Assessment in Software Engineering (EASE)* (Salerno, Italy, 2024), ACM.

[8] DÖDERLEIN, J.-B., ACHER, M., KHELLADI, D. E., AND COMBEMALE, B. Piloting copilot and codex: Hot temperature, cold prompts, or black magic? *SSRN Electronic Journal* (2023).

[9] ET., M. C. Evaluating large language models trained on code. https://github.com/openai/human-eval, 2021. OpenAI HumanEval Benchmark.

[10] ET AL., C. Evaluating large language models trained on code, 2021. OpenAI Codex.

[11] FENG, Z., GUO, D., TANG, D., DUAN, N., LIU, X., GONG, M., SHOU, L., ZHOU, M., YIN, J., AND JIANG, D. Codebert: A pre-trained model for programming and natural languages. In *Proceedings of the 2020 Conference on Empirical Methods in Natural Language Processing (EMNLP)* (2020), Association for Computational Linguistics, pp. 1536–1547.

[12] GARG, S., MOGHADDAM, R. Z., CLEMENT, C. B., SUNDARESAN, N., AND WU, C. Deepperf: A deep learning-based approach for improving software performance, 2022.

[13] JESSE, K., AHMED, T., DEVANBU, P. T., AND MORGAN, E. Large language models and simple, stupid bugs, 2023.

[14] KHOJAH, R., MOHAMAD, M., LEITNER, P., AND GOMES DE OLIVEIRA NETO, F. Beyond code generation: An observational study of chatgpt usage in software engineering practice. *Proceedings of the ACM on Software Engineering 1*, FSE (2024), Article 81.

[15] LEETCODE. Leetcode: The world's leading online programming learning platform. https://leetcode.com/, 2024. Accessed: 2024-11-06.

[16] LIU, J., XIA, C. S., WANG, Y., AND ZHANG, L. Is your code generated by chatgpt really correct? rigorous evaluation of large language models for code generation, 2023.

[17] LIU, Y., ET AL. Ai software reliability: Concepts and related domains. In *2023 IEEE International Conference on Artificial Intelligence and Industrial Applications (AIIIP)* (2023), pp. 287–292.

[18] LU, S., GUO, D., REN, S., SVYATKOVSKIY, A., BLANCO, A., CLEMENT, C. B., DRAIN, D., BING, L., ZHOU, M., AND SUNDARESAN, N. Codexglue: A benchmark dataset and open challenge for code intelligence. *Proceedings of the 2021 Conference on Empirical Methods in Natural Language Processing (EMNLP)* (2021), CodeXGLUE.

[19] NASCIMENTO, N., ALENCAR, P., AND COWAN, D. Artificial intelligence versus software engineers: An evidence-based assessment focusing on non-functional requirements. *University of Waterloo* (2023).

[20] NAVEED, H., AHMAD, A., AHMAD, M., TANVEER, W., JAVED, S. H., SHAIKH, A., MUMTAZ, A., ASLAM, A., AND ILYAS, A. A comprehensive overview of large language models. *arXiv preprint arXiv:2307.06435* (2023).

[21] NGUYEN, N., AND NADI, S. An empirical evaluation of github copilot's code suggestions. In *19th International Conference on Mining Software Repositories (MSR)* (Pittsburgh, PA, USA, 2022), ACM.

[22] OPENAI. Gpt-4 technical report, 2024.

[23] QIHAO ZHU ET AL., YEAR=2024, E. A. P. U. Deepseek-coder-v2: Breaking the barrier of closed-source models in code intelligence.

[24] SOBANIA, D., BRIESCH, M., AND ROTHLAUF, F. Choose your programming copilot: A comparison of the program synthesis performance of github copilot and genetic programming. In *Proceedings of the Genetic and Evolutionary Computation Conference* (2022), pp. 1019–1027.

[25] STRATASCRATCH. Stratascratch: Real-world data science coding challenges. https://www.stratascratch.com/, 2022. Benchmark platform for data science and coding challenges.

[26] SVYATKOVSKIY, A., DENG, S. K., FU, S., AND SUNDARESAN, N. Intellicode compose: Code generation using transformer models. *Proceedings of the 28th ACM SIGKDD Conference on Knowledge Discovery and Data Mining* (2022), 3036–3046.

[27] VAITHILINGAM, P., YANG, G., AND ZHU, H. Expectation vs. experience: Evaluating the usability of code generation tools powered by large language models. In *Proceedings of the 2022 CHI Conference on Human Factors in Computing Systems* (New York, NY, USA, 2022), Association for Computing Machinery, pp. 1–13.

[28] VASCONCELOS, H., BANSAL, G., FOURNEY, A., LIAO, Q. V., AND VAUGHAN, J. W. Generation probabilities are not enough: Uncertainty highlighting in ai code completions. *ACM Transactions on Computer-Human Interaction* (Oct. 2024).

[29] WHITE, J., FU, Q., HAYS, S., SANDBORN, M., SPENCER-SMITH, J., AND SCHMIDT, D. C. Chatgpt prompt patterns for improving code quality, refactoring, requirements elicitation, and software design. *arXiv preprint arXiv:2302.11382* (2023).

[30] XU, F. F., ALON, U., AND NEUBIG, G. Polycoder: A language model for code completion. *arXiv preprint arXiv:2202.13169* (2022).